\documentclass[a4paper,pra,twocolumn,superscriptaddress]{revtex4-1}
\usepackage[english]{babel}
\usepackage[utf8]{inputenc}
\usepackage{graphicx,epsfig,color}
\usepackage{latexsym}
\usepackage{amsfonts} 
\usepackage{amsmath}
\usepackage{textcomp}
\usepackage{hyperref}
\usepackage{placeins}
\usepackage{soul}
\usepackage{cancel}
\usepackage{braket}
\usepackage{subcaption}

\def\<{\langle}
\def\>{\rangle}
\def\{{\lbrace}
\def\}{\rbrace}
\def\({\left(}
\def\){\right)}
\def\beq{\begin{equation}}
\def\eeq{\end{equation}}

\def\ve{\varepsilon}
\def\SS{{\mathcal S}}
\def\RR{{\mathcal R}}
\def\II{{\mathcal I}}

\begin{document}

\title{Effects of confinement and vaccination on an epidemic
  outburst: a statistical mechanics approach}

\author{Óscar Toledano}
\affiliation{Departamento de Física Interdisciplinar, Facultad de Ciencias, UNED, Madrid (Spain), 28040}

\author{Begoña Mula}
\affiliation{Departamento de Física Fundamental, Facultad de Ciencias, UNED, Madrid (Spain), 28040}

\author{Silvia N. Santalla}
\affiliation{Departamento de Física \&\ Grupo Interdisciplinar de Sistemas
  Complejos, Universidad Carlos III de Madrid, Leganés (Spain), 28911}

\author{Javier Rodríguez-Laguna}
\affiliation{Departamento de Física Fundamental, Facultad de Ciencias, UNED, Madrid (Spain), 28040}

\author{Óscar Gálvez}
\affiliation{Departamento de Física Interdisciplinar, Facultad de Ciencias, UNED, Madrid (Spain), 28040}


\begin{abstract}
  This work describes a simple agent model for the spread of an
  epidemic outburst, with special emphasis on mobility and
  geographical considerations, which we characterize via statistical
  mechanics and numerical simulations. As the mobility is decreased, a
  percolation phase transition is found separating a free-propagation
  phase in which the outburst spreads without finding spatial barriers
  and a localized phase in which the outburst dies off. Interestingly,
  the number of infected agents is subject to maximal fluctuations at
  the transition point, building upon the unpredictability of the
  evolution of an epidemic outburst. Our model also lends itself to
  testing vaccination schedules. Indeed, it has been suggested that if
  a vaccine is available but scarce it is convenient to
  carefully select the vaccination program to maximize the chances of halting
  the outburst. We discuss and evaluate several schemes, with special
  interest on how the percolation transition point can be shifted,
  allowing for higher mobility without epidemiological impact.
\end{abstract}

\date{July 24, 2021}

\maketitle

\section{Introduction}

Epidemic containment has been a crucial problem for humankind
throughout our history, which has become of paramount importance since
the COVID-19 outbreak late in 2019. The efforts to understand the
spread of infectious diseases have attracted a large variety of
professionals from all scientific fields, ranging from biology to
sociology (see recent studies from different fields:
\cite{Ben-Hu_2020, Wynants_2020, Thomas_2020}).  Mathematical modeling
has also provided very relevant tools to analyze the stream of data
concerning the infected population, and has substantially contributed
to policy design (see, e.g., \cite{Li_2018, Hebert-Dufresne_2013,
  Riley_2015}).  Among the different approximations employed,
agent-based models have been extensively used to provide policy
recommendations even in the COVID-19 case \cite{Aleta_20}, for which
simulations based on stratified population dynamics were carried out
\cite{Arenas_20}. Multiscale approaches are known to improve our
ability to explain the geographical expansion of the disease, and they
have been also used, along with data-driven simulations, to analyze
the epidemic of COVID-19 in Brazil \cite{Costa_Cota_20}. Epidemic
waves have also been considered \cite{Costa_Cota_20,Munoz_21},
employing stratified population dynamics and non-autonomous dynamics,
where mitigation effects are subsequently imposed and relaxed.

During the COVID-19 epidemic and due to the scarce amount of
vaccination doses in the first stages of the vaccination campaign,
immunization schedules have also attracted attention of the modeling
community with the aim of stifling the expansion of an epidemic burst
by acting upon a number of individuals substantially below the
percolation threshold \cite{Costa_Ferreira_20}, e.g., through the
search of certain types of motifs in the contact network.

However, predictions about the evolution of an epidemic burst are
known to be difficult and unreliable, because the uncertainty in the
initial data propagates exponentially \cite{Castro_20}. Yet, on
occasions the inherently unpredictability of the models can be turned
in our favor.As mentioned in previous works, the fluctuations in the number of
infected people during the evolution of an epidemic burst can provide
useful information regarding our ability to stifle an ongoing
epidemic \cite{Shu_15}.

In this work, we propose a very simple agent model \cite{CSM} of the
susceptible-infected-recovered (SIR) type
\cite{Kermack_1927,Li_book}. Agents follow a random walk in the
vicinity of their homes, which are randomly distributed on a square
lattice, wandering up to a maximum distance $r$, which may vary from
agent to agent. When $r$ is very short, agents are effectively
confined at home, and an infectious outbreak will very likely die
off. As this distance is increased, outbreaks have larger chances of
spreading throughout the system, until a percolation phase transition
is reached, \cite{Stauffer_2003} in which the presence of an infinite
cluster becomes certain. Further increases of mobility will have a
very limited impact on the spread of the infectious
disease. Interestingly, predictions about the future evolution of the
epidemic are more difficult near the transition point. Indeed, above
the percolation threshold the mean-field theory associated with the
SIR equations provide a very accurate prediction of the evolution of
the model, while below the transition point typical outbursts will not
propagate beyond a certain correlation length. But as the agent
mobility approaches to the percolation value, the precise geographical
origin of the outbreak becomes crucial to predict the outcome. Even
though an infinite connected cluster exists, the probability that the
initial infected agent will be part of it is minimal at that point,
leading to a maximal uncertainty. Thus, we show that the fluctuations
in the number of infected people become a very useful observable in
order to pinpoint the phase transition.

Of course, our model is far too simple to be taken decisively in order
to provide policy recommendations, which should be always undertaken
with the assessment of experts from several fields. Yet, our results
suggest that lock-down measures increase their effectiveness very
sharply once the percolation threshold has been crossed. Finding the
minimal lock-down measures that lead to the phase transition is
therefore of paramount importance, and a very difficult task. We can
not provide a complete recipe to establish the location of that
threshold in practice, but we provide an interesting proxy as was noted before \cite{Shu_15}:
fluctuations in the infection reach.

If an effective vaccine is available but scarce, a vaccination
schedule becomes unavoidable, i.e.: the sanitary authorities should
consider which agents must be immunized first. According to their social role, some of these agents can be selected in order to minimize the
spread of the infection \cite{Schneider_2011,Matamalas_2018,Yang_19}. Of course, many individuals should receive
the vaccine because of other considerations, such as age or health
condition, and we will not discuss these very relevant
aspects. Our model lends itself very easily to the evaluation of the
efficiency of a vaccination program. In it, individuals are only
distinguished through their relations: some homes are relatively
isolated, so their inhabitants are not likely to spread the
disease. Yet, a naive approach would consider that individuals with
many connections should be the first candidates to be immunized. We
will show that this criterion is not optimal. Indeed, some agents
with few connections act as natural {\em bridges} between different
clusters. Thus if they receive the vaccine the clusters will become
isolated and the outburst will be halted.

This article is organized as follows. Section \ref{sec:model} discusses
in detail our agent model, and how our numerical simulations are
performed. In Section \ref{sec:simulations} we describe our results
regarding the percolation phase transition as the agent mobility is
decreased, the secondary cases and the effects of a finite probability
of recovery. Our theoretical framework is discussed in
Section \ref{sec:theory}. The efficiency of several vaccination schemes
is discussed in Section \ref{sec:vaccination}, especially in connection
with the shift of the percolation threshold. Our conclusions and ideas
for future work are discussed in Section \ref{sec:conclusions}.


\section{The confined-SIR model}
\label{sec:model}

We propose a simple model to characterize the effects of partial
confinement during an epidemic expansion, which we call the {\em
  confined-SIR} model.

Let us consider $N$ agents moving on an $L\times L$ square lattice
with periodic boundary conditions. Agent $i$ possesses a {\em home},
determined by a fixed lattice point, $\vec H_i$. Agents can move
freely within their {\em wandering circles}, centered at $\vec H_i$
and with radius $r_i$, which we will assume equal, $r_i=r$ for all
$i$. Time advances in discrete steps $\Delta t=1$, in arbitrary
units. At each time step, every agent performs a random step.
If the step takes the agent out of the wandering circle a
new step is attempted. Homes are distributed randomly over the whole
lattice.

\begin{figure}
  \includegraphics[width=7.5cm]{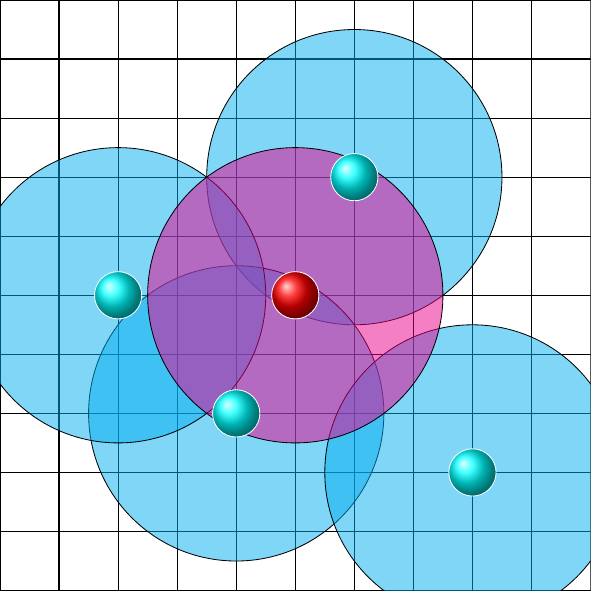}
  \caption{Illustration of the confined SIR model. The red agent at
    the center is infected and can move freely within its wandering
    circle, shown in magenta. Blue agents are susceptible and can move
    within their wandering circles, shown in cyan. Notice that the
    magenta circle intersects all cyan circles, thus providing a
    finite probability of infection for all those agents.}
  \label{fig:illust}
\end{figure}

Agents can be classified into three compartments: susceptible ($S$),
infected ($I$) and recovered ($R$). Susceptible agents get infected with probability $n\beta$ when they share cell with $n$ infected agents.
Infected agents may get recovered at each time-step, with
probability $\gamma$. In some cases, an effective SIR model can be
written down and solved analytically. An illustration is provided in
Fig. \ref{fig:illust}. 

\bigskip

Let $A_i$ stand for the number of lattice points in the wandering
circle of agent $i$, while $A_{ij}$ will stand for the number of
lattice points in the intersection between the wandering circles of
agents $i$ and $j$. The probability that agent $i$ and agent $j$ will
collide at time $t$ can be estimated as

\begin{equation}
  C_{ij}={A_{ij}\over A_i A_j}.
  \label{eq:collision_prob}
\end{equation}
Thus, if agent $i\in S$ and $j\in I$, the probability per unit time
that agent $i$ will get infected by $j$ becomes $P^I_{ij}=\beta
C_{ij}$. Thus, we can find the total probability per unit time that an
agent $i$ will get infected by just summing over all possible infection
sources

\begin{equation}
  P^I_i = \beta  \sum_{j\in I} C_{ij} = \beta \sum_{j\in I} {A_{ij}\over A_i A_j}.
  \label{eq:infection_prob}
\end{equation}

We will be mostly interested in the continuum limit, in which
$L\to\infty$, $N\to\infty$ but the density $\rho=N/L^2$ is constant,
in order to avoid lattice effects. In that limit, $A_i\approx \pi r^2$
and $A_{ij}$ corresponds to the overlap between two circles whose
centers are a distance $d_{ij}$ apart, which is given by

\begin{equation}
  A_{ij}\approx 2r^2 \(\arccos(u_{ij})-u_{ij}\sqrt{1-u_{ij}^2}\)
  \equiv 2r^2 f(u_{ij}),
  \label{eq:intersect}
\end{equation}
where $u_{ij}=d_{ij}/(2r)\in [0,1]$ is the dimensionless relative
distance. Thus, we have

\begin{equation}
P^I_i \approx \beta \sum_{j\in I} {2f(u_{ij}) \over \pi^2 r^2}
\approx \beta\rho {2\over\pi^2} \left[\int_0^1 du\; u f(u)\right]
\equiv K\beta\rho,
\label{eq:PI}
\end{equation}
showing that the effective infection probability does not depend on
the wandering radius $r$.

\bigskip

The dynamics of the model can be approximated by a Markov chain. Let
us consider that, at time $t$, we know the epidemiological status of
each agent, the location of their home $\vec H_i$, and the wandering
radius $r$. The evolution of the expected number of infected agents
can now be found,

\begin{equation}
  \<I(t+1)\> = \(1-\gamma\) \<I(t)\>+ \beta
  \left\langle \sum_{\substack{i\in S\\j\in I}} C_{ij}\right\rangle,
  \label{eq:Ievol}
\end{equation}
Notice that the $C_{ij}$ are constants, but the right-hand side
evolves due to the migration of agents between sets $I$ and $S$. In
the totally mixing limit, $r\sim L$ and all agents wander around the
whole region. Therefore, all $C_{ij}\sim 1/L^2$. Thus, in a mean-field
approximation Eq. \eqref{eq:Ievol} becomes

\begin{equation}
  \<I(t+1)\> \approx (1-\gamma) \<I(t)\> + {\beta \over L^2}
  \<S(t)\>\<I(t)\>,
  \label{eq:meanfield}
\end{equation}
and we obtain the usual SIR differential equations. Introducing
fractional variables, $\SS=\<S\>/N$, $\II=\<I\>/N$ and
$\RR=\<R\>/N$ we reach:

\begin{align}
\dot\SS &\approx -\beta \rho\; \SS \II, \nonumber\\
\dot\II &\approx -\gamma\; \II + \beta\rho\; \SS(t) \II,\nonumber\\
\dot\RR &\approx \gamma\; \II.
\label{eq:meanfield2}
\end{align}
For early times, $\SS\approx 1$ so the infection either expands or
vanishes depending on the value of the basic reproductive number,
which in this case can be defined as

\begin{equation}
  R_0\equiv {\beta\rho \over \gamma}.
  \label{eq:R0}
\end{equation}

Our main focus will be the probability distribution for the total size of the outburst, characterized by the number of affected agents, $A$, that have suffered the infection at any moment in the long run, and its deviation, $\sigma_A$.
Of course,
there are other relevant observables, such as the maximum number of
infected agents, but they are not be considered in this work.

\bigskip

\begin{figure}
  \includegraphics[width=8cm]{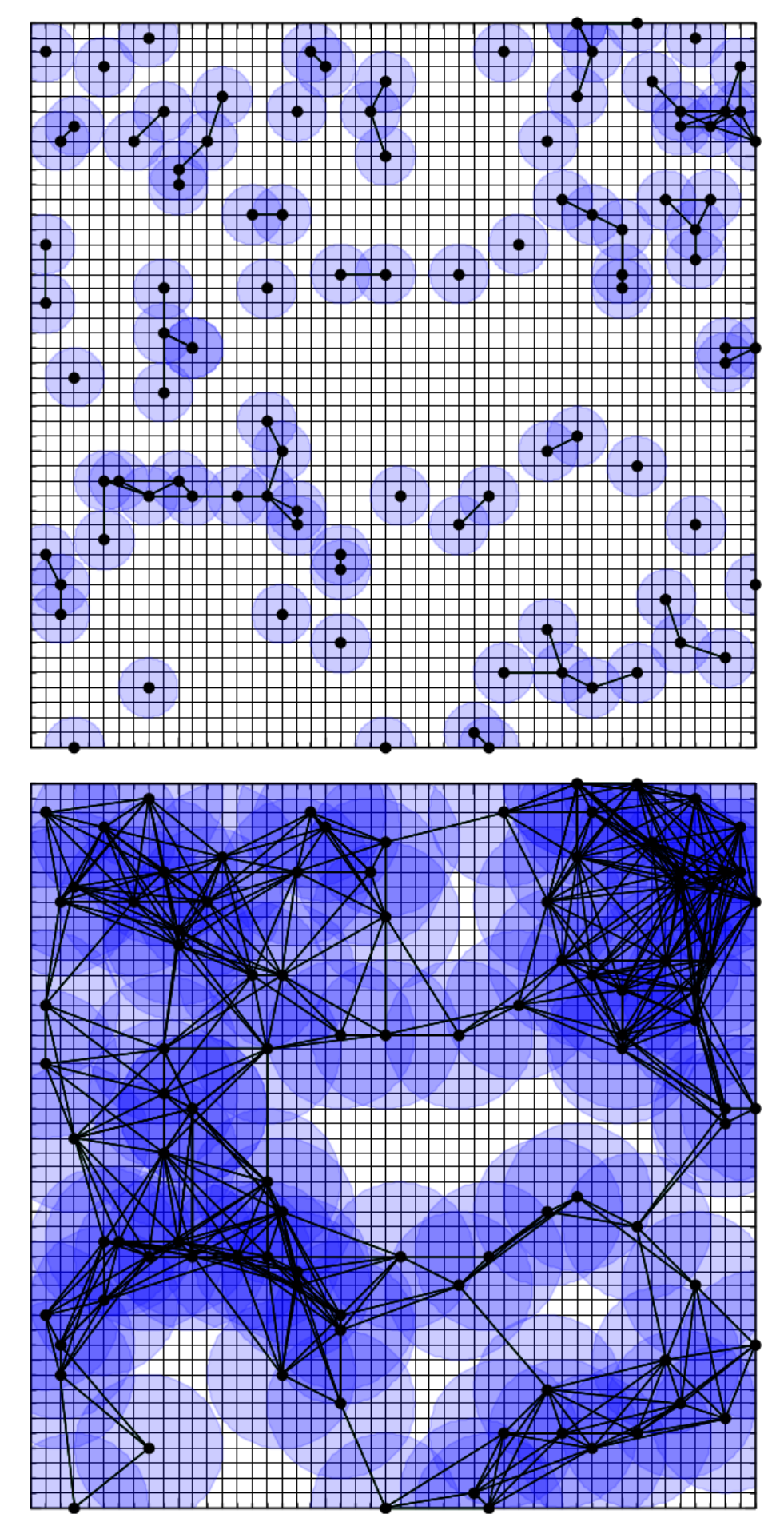}
  \caption{Effective networks obtained from a confined-SIR system
    using $N=100$ agents on a $50\times 50$ lattice. Nodes are
    associated with the homes of the agents, while links are drawn if
    their respective dwellers are able to meet during their
    wandering. Top: using $r=2$; bottom: using $r=5$.}
  \label{fig:effective}
\end{figure}

Our model presents a very clear graph structure. Indeed, two agents
$i$ and $j$ may infect each other if and only if $C_{ij}>0$. This
condition determines an effective graph ${\cal G}$, such as the one
shown in Fig. \ref{fig:effective}, in which the homes are represented
as nodes and the links denote the pairs of agents that are able to
meet. Both networks are obtained for $N=100$ agents on $50 \times 50$
lattices, using $r=2$ (top) and $r=5$ (bottom). It can be readily seen
that the network for $r=2$ contains many disconnected clusters, while
for $r=5$ it contains a large cluster spanning most of the nodes.

These networks are examples of {\em random geometric graphs} (RGGs)
\cite{Penrose_book,Penrose_1997,Dall_02,Diaz_09}, in which $N$ points
are randomly scattered on a $L\times L$ square and joined with edges
whenever their distance is less than $r$. RGG have proved to be a very
convenient framework for the analysis of epidemic outbursts
\cite{Estrada_2016}, typically providing infected nodes with a certain
probability per unit time of extending the infection to their
neighbors. Yet, our confined-SIR model presents a substantial
difference with those models. Indeed, the graph structure is
equivalent to that of an RGG model, but the actual distances are still
relevant in our case. Since our infected agents perform a random walk,
their collisions into susceptible agents typically take place one by
one, instead of through multiple infections. This difference has been
shown to be relevant in the literature \cite{Gomez_10}.


\section{Numerical simulations}
\label{sec:simulations}

We have run numerical simulations of our confined-SIR model on a
square lattice, placing $N$ random homes and using the same value of
the wandering radius $r$ for all agents. Unless otherwise specified,
we always average over $N_S=5000$ samples, with fixed values for
$\beta=1$ and $\gamma=0$, $N=1000$ agents, and $\rho=5 \cdot
10^{-6}$.

\begin{figure}
  \includegraphics[width=8cm]{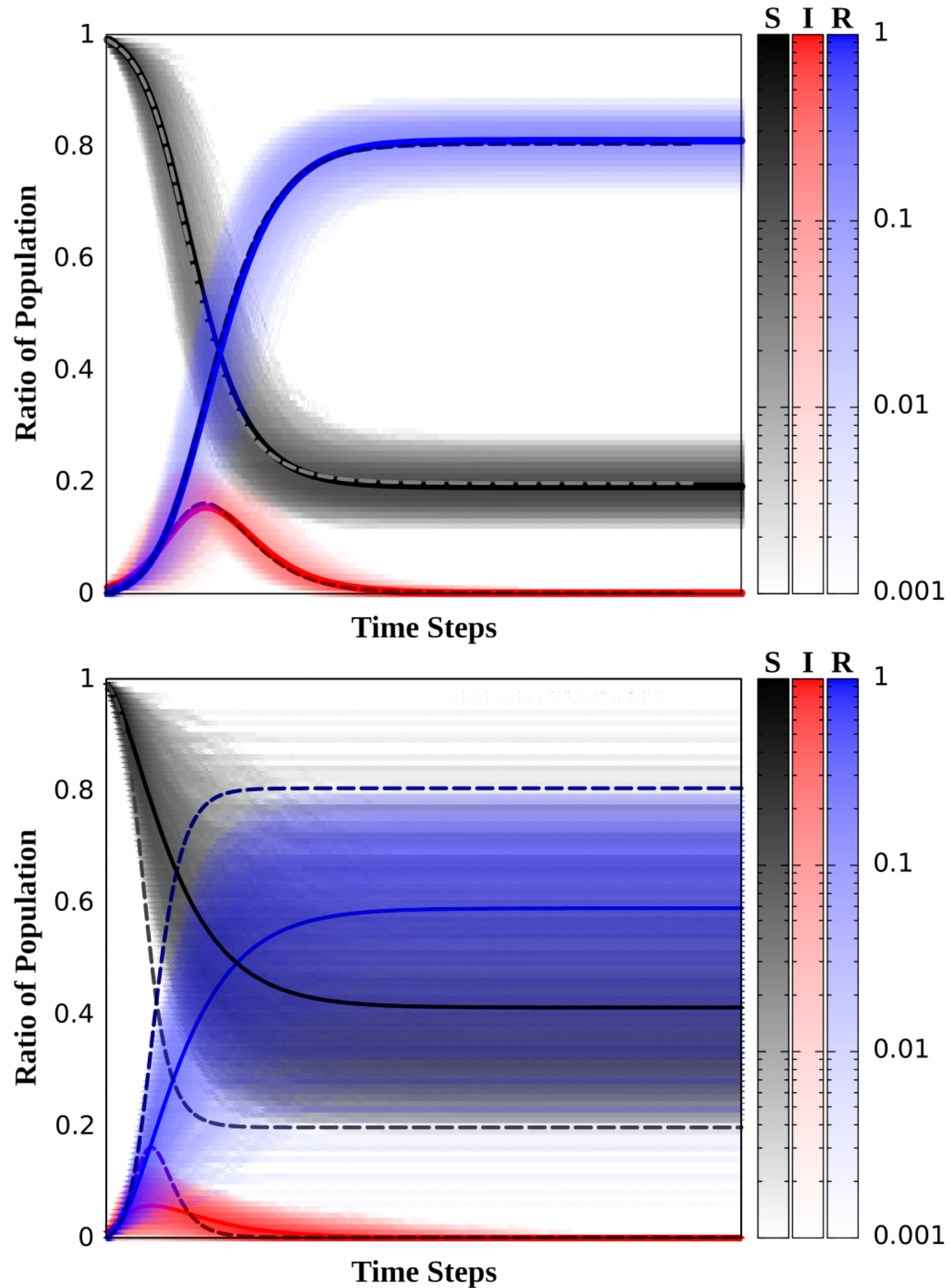}
  \caption{Fraction of susceptible, infected, and recovered agents as a
    function of time, depicted with continuous black, red, and blue
    lines, respectively, shown as a time-dependent histogram through
    the color gradation (more intense corresponding to larger
    probabilities), along with their average values shown in
    continuous lines and the theoretical predictions of the SIR model
    in dashed lines. In both panels we have used $\beta=0.02$,
    $\gamma=5\cdot 10^{-4}$, $L=141$ and $N=1000$. The top panel shows
    the results for a wandering radius $r=40.3$ and the lower panel for
    $r=8.9$.}
    \label{fig:sir}
\end{figure}

For large values of the wandering radius $r$ we expect the epidemic
burst to follow the mean-field approximation provided by the SIR
equations for perfect mixing, Eq. \eqref{eq:meanfield2}. We have
checked that conjecture numerically in Fig. \ref{fig:sir} (top), where
we show the probability distribution for the fractions $\SS$, $\II$
and $\RR$ as a function of time, along with their expected values for the
theoretical prediction. However, the results for smaller values of $r$
are very different, as we can see in Fig. \ref{fig:sir}
(bottom). Indeed, in this case the average values of $\SS$, $\II$ and
$\RR$ differ notably from the theoretical predictions, because the
mixing assumption is inadequate in this regime.

Thus, we are led to identify two different regimes. For large values
of $r$ we have nearly perfect mixing and for low $r$ values the
system is naturally divided into isolated clusters. Both are separated
by a {\em percolation transition}, which we characterize in the next
section.

\subsection{Percolation phase transition}

The theory of bond percolation has been one of the foremost paradigms
of statistical mechanics for more than 40 years
\cite{Hammersley_65,Howard_04,Auffinger_17,Cordoba_18}, with
applications to magnetism \cite{Abraham_95}, wireless communications
\cite{Beyme_14}, ecological competition \cite{Kordzhakia_05} or
sequence alignment in molecular biology \cite{Bundschuh_00}. Recently,
a very relevant connection was described between the geodesics in
strongly disordered networks and bond percolation
\cite{Villarrubia.20}. Application of percolation theory to epidemics
has been carried out by previous authors, such as those of Refs. \cite{Sander_2003,
  Miller_2009}, or more recently, Refs. \cite{Oliveira_2020, Croccolo_2020}.

Bond percolation on a fixed lattice is characterized by a single
parameter, $p$, the probability that a given bond will be
present. Above a certain threshold value, $p>p_c$, the probability
that the system will contain an infinite connected cluster reaches
one, with $p_c=1/2$ for the square lattice. Our system presents a
strong similarity with bond percolation, but with another observable
playing the role of the bond probability $p$. Let us consider $N$
agents on an $L\times L$ lattice. The average distance between homes,
$\bar r$, can be estimated as

\beq
\bar r={L\over\sqrt{N}}.
\label{eq:r_bar}
\eeq
Now, let us notice that the wandering radius $r$ is only meaningful
when compared to this average distance between homes. Thus, we
introduce a {\em mobility parameter},

\beq
\ve={r\over \bar r}={r\sqrt{N}\over L}.
\label{eq:rho}
\eeq
We will readily show that the mobility parameter $\ve$ is the only
relevant variable to determine the geometry of our system.

\begin{figure}
  \includegraphics[width=8cm]{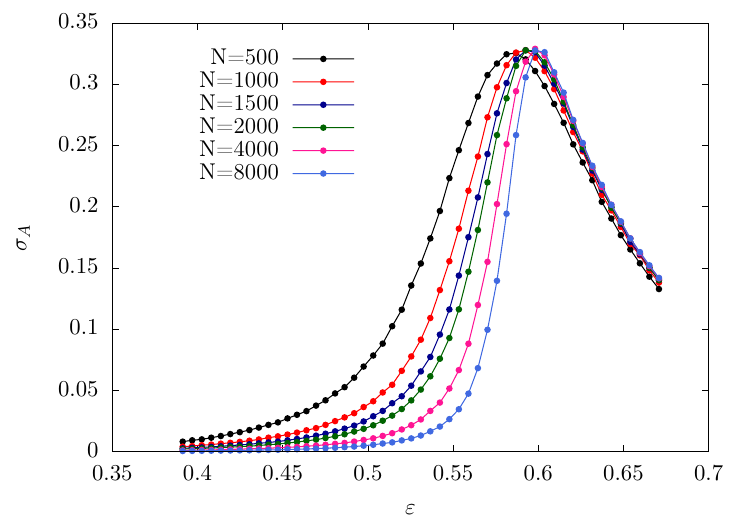}
  \includegraphics[width=8cm]{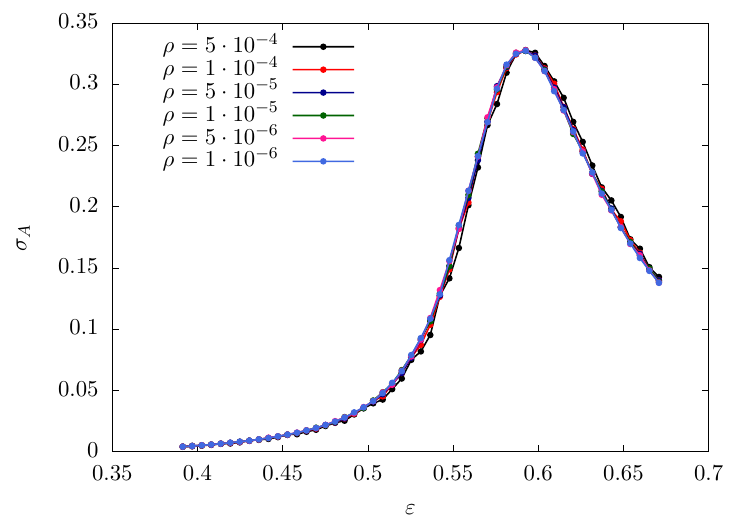}
  \caption{Top: Deviation of the long-term fraction of affected
    agents, $\sigma_A$, as a function of
    the mobility parameter $\ve$, for different numbers of agents and
    a fixed density $\rho=N/L^2=5\cdot 10^{-6}$, $\beta=1$ and
    $\gamma=0$. Bottom: Same observable as a function of the mobility
    parameter $\ve$ using $N=1000$ agents for several values of the
    agent density $\rho$. Notice the collapse of all curves.}
  \label{fig:fluct}
\end{figure}

Critical points, such as the percolation transition, typically lead to large fluctuations, and can be obtained by considering the deviation of the number of affected agents, $\sigma_A$. This magnitude is shown in
Fig. \ref{fig:fluct} as a function of the mobility parameter for
different lattice sizes (top) and densities (bottom). The validity of this strategy for characterizing the transition point has already been probed in the literature
\cite{Ramirez_12,Shu_15}. Indeed, we can observe that the fluctuations
in the size of the outburst present a maximum for a certain value of
$\ve$ which only depends on the ratio between the recovery and
infection probabilities, $\gamma/\beta$.

\begin{figure}
  \includegraphics[width=8cm]{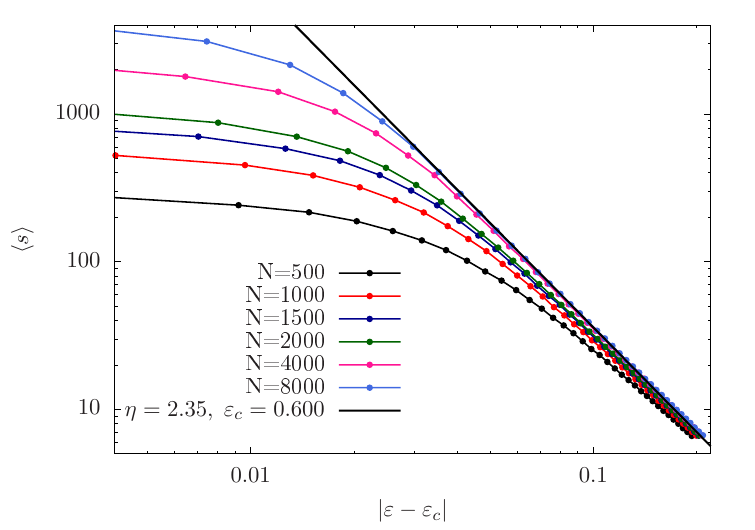}
  \caption{Average cluster size as a function of $\ve-\ve_c$, using
    $\rho=5\cdot 10^{-6}$, $\beta=1$ and $\gamma=0$. Notice the
    power-law behavior, following Eq. \eqref{eq:s_average}, with an
    exponent $\eta\approx 2.35$, close to the theoretical prediction
    $\eta=43/18$.}
  \label{fig:clustersize}
\end{figure}

Let us provide evidence that this critical value of the mobility
parameter, $\ve_c$, corresponds to the position of the percolation
phase transition. In the vicinity of the percolation phase-transition
many observables show critical behavior in the form of power laws. The
most salient of those is given by the average size of a cluster,
$s$. Below the transition, we have

\beq
\braket{s} \sim |p-p_c|^{-\eta},
\label{eq:s_average}
\eeq
with $\eta=43/18\approx 2.39$ for a square lattice
\cite{Stauffer_2003}. Figure \ref{fig:clustersize} shows the average
cluster size of our effective networks as a function of the mobility
parameter $\ve$, for different system sizes and populations. We can
observe that, for all system sizes considered, the average size of the
cluster diverges as we approach a critical value $\ve_c$, with an
exponent which slightly differs from the value obtained in the square
lattice, $\eta\approx 2.35$, which seems to be robust under changes in
the lattice size and the density.

\subsection{Secondary cases}

Epidemic phenomena are characterized by the number of secondary cases,
defined as the number of susceptible agents infected by a single
infected node. The average value of this number is related to $R_0$,
defined in Eq. \eqref{eq:R0}. Yet, the deviation of that number has
also proved to be a valuable tool to understand the evolution of an
epidemic outburst. Indeed, given the exponential nature of the
expansion, large fluctuations in the spread rate will naturally
dominate the long-term evolution of the disease
\cite{Adam_20,Lewis_21}.

\begin{figure}
  \includegraphics[width=8cm]{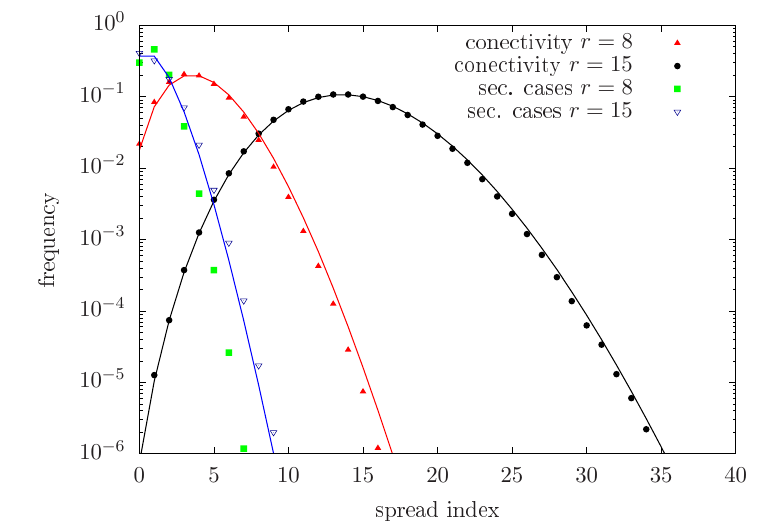}
  \caption{Histogram of secondary cases in the confined-SIR model for
    $N=1000$, $L=447$, using $r=8$ and $r=15$, corresponding to
    $\ve=0.566$ and $\ve=1.060$. For comparison, we
    also show the connectivity histogram in the associated RGG and
    the fit to a Poisson law.}
  \label{fig:spread}
\end{figure}

It is thus relevant to wonder whether the spread of the infection
within our model gives rise to large deviations in the number of
secondary cases, which we will characterize using its full
histogram. In RGG models, a proxy for that value is the number of
connections for each node. Fig. \ref{fig:spread} shows the full
histogram for the number of neighbors of any given node, both before
and after percolation, i.e., for $r=8$ and $15$ using
$N=1000$ and $L=447$. The data fit perfectly
to a Poisson distribution, corresponding to the theoretical
expectation. Furthermore, the figure also shows the number of actual
secondary cases, which is much more peaked, showing that the deviation
is always low.

The reason for this difference between RGG and our model lies in the statistical properties of time sequences of infection events involving a common infected agent. In an RGG setup infection times are uncorrelated Poisson variates, but in our model those times are not independent. Indeed, an infected agent may require a minimal time to infect two different neighbours, given by the relative positions of the overlapping areas between the three wandering circles involved.

\subsection{Non-Zero Recovery Probability}

Let us discuss the effect of a non-zero recovery probability,
$\gamma>0$. In that case, infected agents may recover before they can
propagate the disease, and thus we are led to compare two natural
times: $\tau_{ij}\approx (\beta C_{ij})^{-1}$ is the expected time before the
infection may propagate from agent $i$ to agent $j$ (or viceversa),
while $\tau_R\approx \gamma^{-1}$ corresponds to the expected time
before recovery. Thus, if $\tau_{ij} \gg \tau_R$ we may assume that
the infection will not be able to propagate from agent $i$ to agent
$j$, while in the opposite case, $\tau_R \gg \tau_{ij}$, we may neglect
the possibility of recovery. We may claim that a finite recovery
probability provides an effective cutoff for the local infection
probabilities, thus removing weak links from the graph.

\begin{figure}
  \includegraphics[width=8cm]{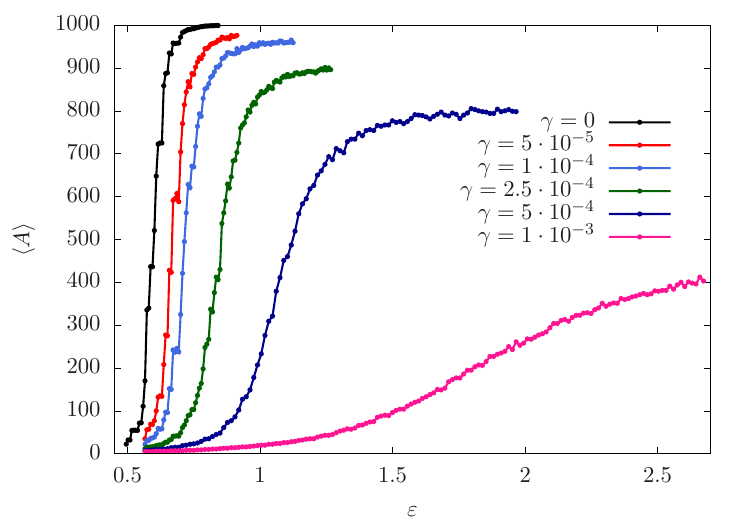}
  \includegraphics[width=8cm]{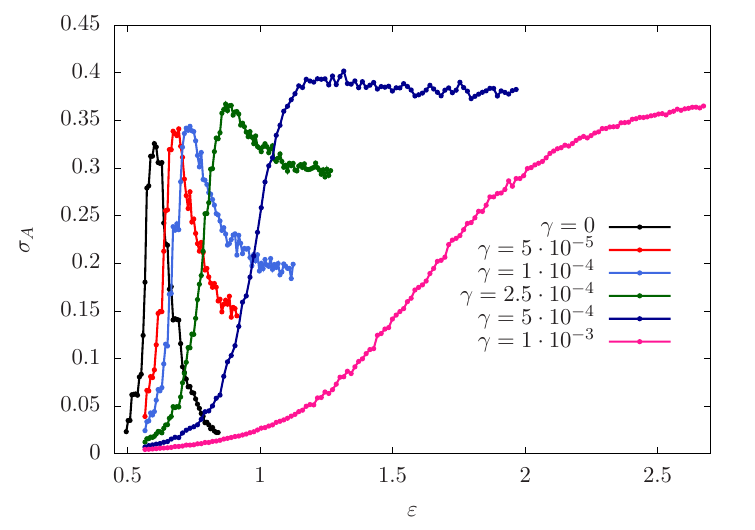}
  \caption{Top: average number of affected agents, $\<A\>$, as a
    function of the mobility parameter for different values of the
    $\gamma$ parameter. Bottom: Deviation of the same magnitude,
    $\sigma_A$. Simulations were performed with $N=1000$, $N_S=5000$
    samples, $\beta=1$, and $\rho=5\cdot 10^{-3}$.}
  \label{fig:fluct_rec}
\end{figure}

The top panel of Fig. \ref{fig:fluct_rec} shows the average number of
affected nodes, $\<A\>$, as a function of the mobility parameter as we
increase the recovery probability per unit time, $\gamma$. It shows a
continuous decrease, as expected. The bottom panel of
Fig. \ref{fig:fluct_rec}, on the other hand, shows the deviation in
the number of affected nodes, $\sigma_A$, showing that the
fluctuations at the maximum do not depend strongly on $\gamma$. In
addition, the value of the mobility parameter for which this maximum
takes place grows rather fast with $\gamma$. Thus we are led to claim
that the recovery probability per unit time strongly affects the value
of the percolation threshold. Also, we can see in the bottom panel of
Fig. \ref{fig:fluct_rec} that for low $\gamma$ the fluctuations show a
peaked maximum, while it seems to reach a plateau for higher values
of $\gamma$.


\section{Theoretical framework}
\label{sec:theory}

Let us provide a theoretical framework in order to explain the nature
of the phase transition observed in the system. A percolation
transition of geometrical origin is combined with an epidemic spread
transition that takes place when $R_0>1$ in a well mixed system. 

\begin{figure}
  \includegraphics[width=8cm]{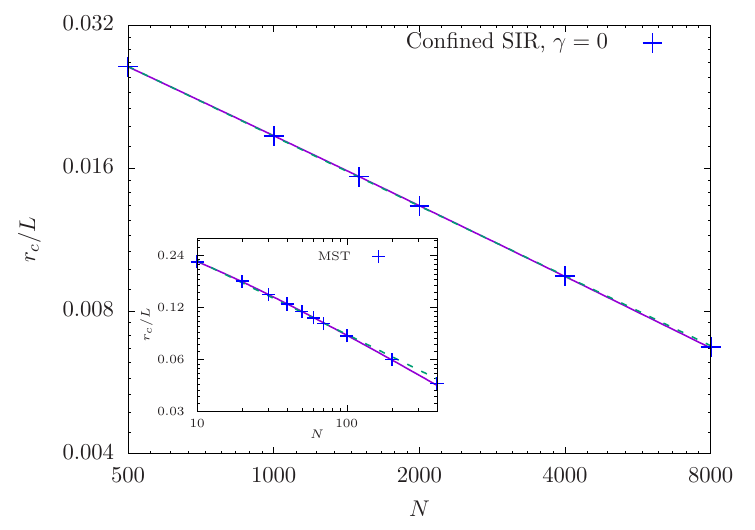}
  \includegraphics[width=8cm]{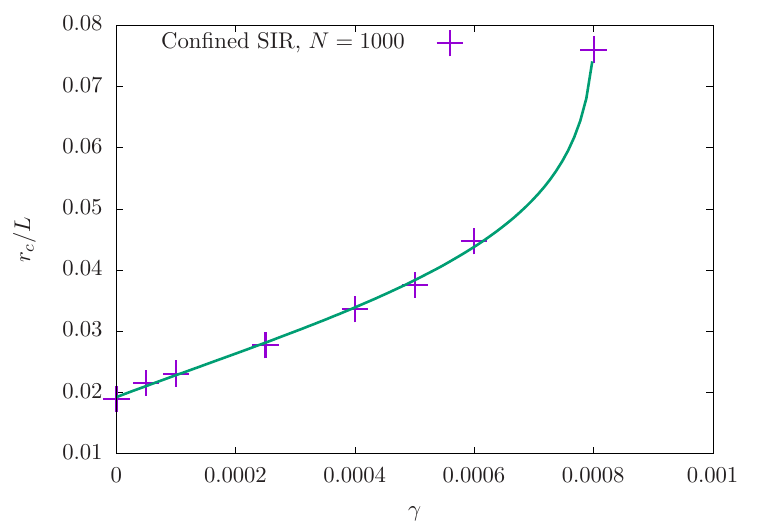}
  \caption{Top: Transition radius as a function of the size system $N$
    for $\gamma=0$. The dashed line is a fit to expression
    \eqref{eq:fit1}, while the slightly more accurate continuous line
    shows a fit to expression \eqref{eq:fit2}. Bottom: Transition
    radius as a function of $\gamma$, for $N=1000$, showing a fit to
    expression \eqref{eq:fit3}.
    The transition value of $r_c / L$ is approximately equivalent to $\<\ell\>$ as shown in \eqref{eq:penrose}.}
  \label{fig:transition}
\end{figure}

First, let us consider the percolation transition on an RGG of size
$L\times L$ with $N$ node and radius $r$. The probability that a
single node will be isolated is $1-\pi (2r/L)^2$
\cite{Penrose_book}. The expected number of isolated nodes will be
approximately given by

\begin{equation}
  \mu \approx N\left[1-\pi \({2r\over L}\)^2\right]^{N-1}\approx
  N\exp(-4\pi \ve^2).
  \label{eq:isolated}
\end{equation}
The average connectivity of a node will be $C=4\pi\ve^2$, thus showing
that the number of isolated nodes will be given by $\mu \sim Ne^{-C}$
\cite{Dall_02}. Interestingly, this magnitude drops quickly as $r$
increases at the percolation transition, for large $N$. More
rigorously, it has been proved that the longest edge of the minimal
spanning tree (MST), $\ell$, scales like

\begin{equation}
 \({r_c\over L}\)^2 \sim \<\ell^2 \> \sim {\log N\over {\pi N}},
  \label{eq:penrose}
\end{equation}
for large $N$ \cite{Penrose_1997}. The longest edge of the MST is a
good approximation for the minimal value of $r$ for which a giant
component will exist within our graph, thus signaling the percolation
transition. The validity of Eq. \eqref{eq:penrose} can be checked in
the top panel of Fig. \ref{fig:transition}, which shows our estimate
for the critical radius $r_c$ when $\gamma=0$ using several values of
$N$, along with two different fits,

\begin{equation}
  {r_c\over L}=\sqrt{K\over N^\chi},
  \label{eq:fit1}
\end{equation}
which yields an exponent $\chi=1.1$ or the slightly more
accurate fit to

\begin{equation}
  {r_c\over L}=\sqrt{K {\log(N)\over N^\chi}},
  \label{eq:fit2}
\end{equation}
which yields $\chi=0.98$. The inset shows the results for the
longest edge of the MST of graphs of the same type, along with both
types of fits. 

\bigskip

Now, let us shift our attention to the characterization of the
epidemic spread transition on generic networks
\cite{Wang_17,Estrada_2016}. The standard approach to locate the
spread transition on a random graph is to equate the expected number
of infections and the expected number of recoveries in the unit time,
which is usually expressed in the mean-field equation,

\begin{equation}
  \tilde\beta \<k\> = \gamma,
  \label{eq:bkgamma}
\end{equation}
where $\<k\>$ is the average degree of the network and
$\tilde\beta=K\beta\rho$ is the average infection rate for neighboring
nodes, given by Eq. \eqref{eq:PI}. It has been shown that $\<k\>$ can
be replaced with the largest eigenvalue of the adjacency matrix
$\Lambda_{\text{max}}$, providing better results
\cite{Preciado_10}. Even more, we may take into account the presence
of correlations \cite{Silva_2020}. Yet, Eq. \eqref{eq:bkgamma} is good
enough as a first approximation.

In our case, we will consider $\tilde\beta\<k\>$ to be the average
value of the total infection rate, given by Eq. \eqref{eq:PI}, but
{\em restricting ourselves to nodes belonging to the largest connected
  component}. The probability that a node will belong to this giant
component may be approximated as $p_C\approx 1-\exp(-4\pi \ve^2)$, as we
can readily check from Eq. \eqref{eq:isolated}. Thus, we can write
down a fundamental equation defining the critical mobility $\ve_c$,

\begin{equation}
  \gamma_0 \(1-e^{\kappa_1-(\kappa_2\ve)^2}\) = \gamma,
  \label{eq:transition_1}
\end{equation}
where $\gamma_0=K\rho\beta$, and $\kappa_1$ and $\kappa_2$ are fitting
parameters. Solving for $\ve_c$ we obtain

\begin{equation}
\ve_c = \kappa_2 \sqrt{ \kappa_1 - \log\( {1\over 1-\gamma/\gamma_0} \)},
\label{eq:fit3}
\end{equation}
where $\gamma_0=K\beta\rho$. Equation \eqref{eq:fit3} shows that
$r_c\to\infty$ as $\gamma\to \gamma_0$, showing that the infection
will never spread beyond that point. The bottom panel of
Fig. \ref{fig:transition} shows the critical radius obtained for
$N=1000$ as a function of $\gamma$, along with a fit to
Eq. \eqref{eq:fit3}, with $\kappa_1$, $\kappa_2$ and $\gamma_0$ as fitting
parameters. Interestingly, the fit is able to guess a suitable value
of $\gamma_0\approx 0.0008$, despite the fact that the provided values
were far from it.

\section{Vaccination schedules}
\label{sec:vaccination}

Let us consider the possibility of providing immunity through
vaccination to a (small) fraction of the population, $f_v$, for the
case of zero recovery rate, $\gamma=0$. In this section we will
attempt to answer the following question: How do we select the agents
that will receive the vaccine, if our only aim is to minimize the size
of a future epidemic outburst? Notice that, in practice, many other
issues must be considered in this situation, such as the health
conditions or the age of the patients.

The simplest vaccination schedule is merely to randomly select the
individuals. Of course, we do not expect this schedule to be very
efficient. We have considered several observables which can be
employed to determine how useful it will be to provide the vaccine to
a certain agent. The simplest one is the {\em total degree}, defined
as the number of first neighbors in the effective graph. Naively, we
might sort the agents by their degrees, and immunize the first
$f_vN$. Yet, it is more efficient to {\em recompute} the degree of
each agent after each selection, and we will do so unless otherwise
specified.

The most promising observable is, nonetheless, the {\em
  betweenness-centrality} (BC) \cite{Freeman_1979} associated to each
agent, $BC_i$, defined as the fraction of the total number of
geodesics which go through agent $i$. This measure is global, and
takes $O(N^4)$ steps to compute using Dijkstra's algorithm to evaluate
the geodesics, or worse when we recalculate the $BC_i$ after each
vaccination \cite{Cormen}. In previous studies vaccination schemes
based on immunizing the highest BC links have proved to bet very
efficient (see e.g. \cite{Schneider_2011}). Considering the high
computational cost of the BC, we are led to propose a cheaper
alternative, the {\em local betweenness-centrality} (LBC)
\cite{Mahyar_2018}, which is defined for each site as its BC
corresponding to a subgraph restricted to itself and its nearest
neighbors. In intuitive terms, the LBC is high for a node that
connects neighbors which are otherwise disconnected among themselves,
and is naturally related to the clustering coefficient. 

Thus we have considered the following
five vaccination schedules.

\begin{enumerate} \setcounter{enumi}{-1} 
\item[{(0)}] No vaccination, considered the base case.
\item[{(1)}] Select randomly.
\item[{(2)}] Select the agents with highest {\em degree} (HD).
\item[{(3)}] Select the agents with highest {\em local betweenness-centrality} (LBC).
\item[{(4)}] Select the agents with highest {\em betweenness-centrality} (BC).
\end{enumerate}

\begin{figure*}
  \includegraphics[width=8cm]{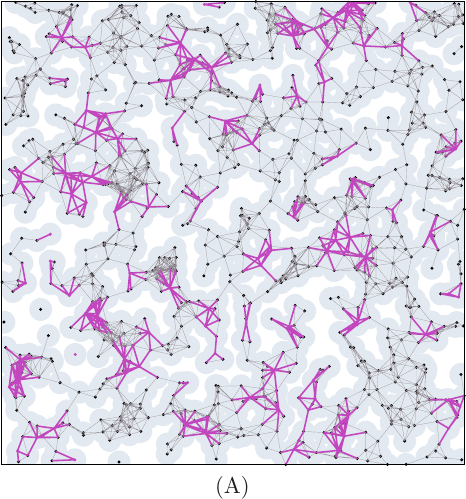}
  \includegraphics[width=8cm]{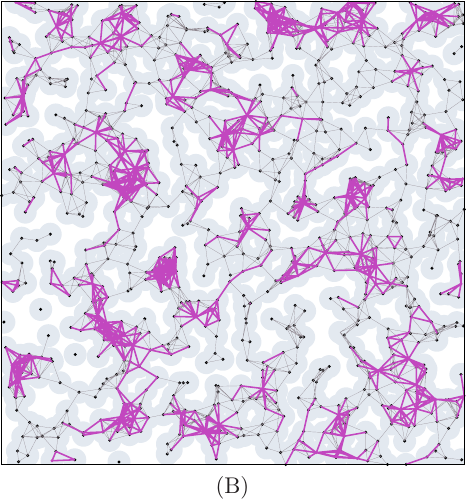}
  \includegraphics[width=8cm]{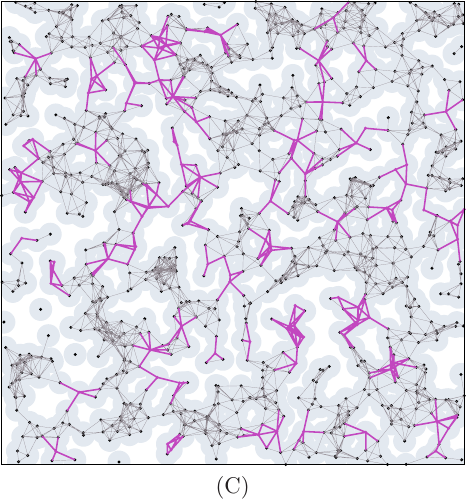}
  \includegraphics[width=8cm]{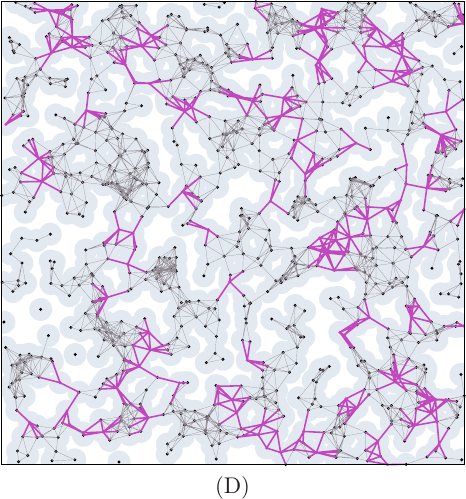}
  \caption{Effect of different vaccination schedules on a fixed
    network: (a) random vaccination, (b) highest degree, (c) highest
    local betweenness-centrality (LBC), (d) highest
    betweenness-centrality (BC). Black dots denote susceptible agents,
    and they are joined by black links. Immunized agents are colored
    purple, same as the dead links. The parameters for all cases
    are $N=1000$, $\varepsilon=0.89$ and a 10\%\ vaccination
    fraction. Videos of the numerical simulations for these vaccinations
    schemes can been seen in \cite{OT_2021}.}
  \label{fig:vacc_networks}
\end{figure*}

The vaccination programs effectively change the topological properties
of the network whenever the removed agents are not selected at
random. Thus, in Fig. \ref{fig:vacc_networks} we can observe a
specific example of a network where different vaccination schemes have
been performed over the same amount of agents. In all cases, let us
focus on the cluster structure. For random vaccination,
Fig. \ref{fig:vacc_networks} (a), the large clusters remain
untouched. For a degree-based vaccination scheme,
Fig. \ref{fig:vacc_networks} (b), we can see that links have been
removed from the {\em core} of the clusters, but the clusters
themselves remain connected. Indeed, immunizing the individuals with a
large number of connections seems to have a low impact on the network
structure in our case. The reason is that high-degree agents tend to
be neighbors of other high-degree agents. Panel (c) of
Fig. \ref{fig:vacc_networks}, on the other hand, shows that removing
agents with a large LBC breaks up some clusters, but some large
clusters still remain active, leading to a likely propagation of an epidemic
outburst to a substantial fraction of the
population. Fig. \ref{fig:vacc_networks} (d) shows the resulting
network when the agents with a largest BC have been removed, and we
can readily see that all large clusters have indeed disappeared.

\begin{figure}[h]
  \includegraphics[width=8cm]{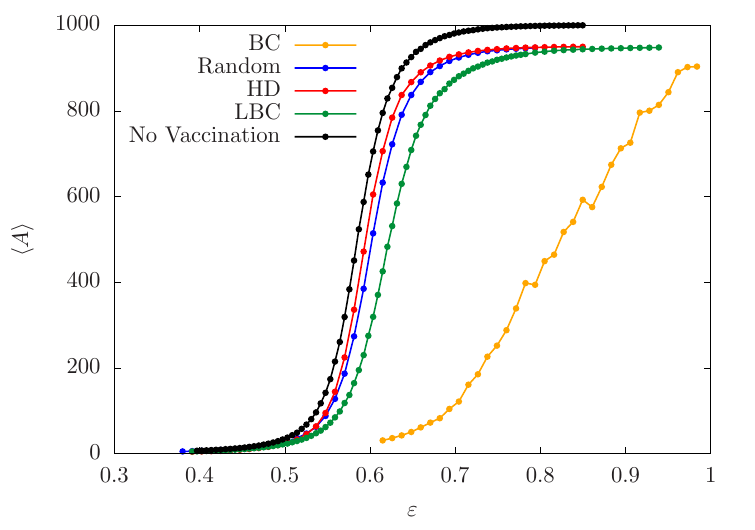}
  \includegraphics[width=8cm]{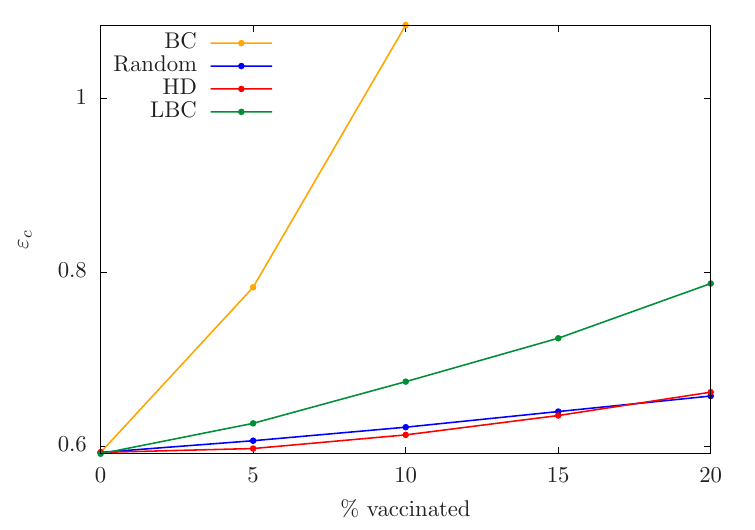}
  \caption{Top: Average number of affected agents for each vaccination
    schedule as a function of the mobility parameter $\ve$ with a
    vaccination fraction of 5\%. Bottom: Mobility parameter marking
    the percolation phase transition, $\ve_c$, for each vaccination
    schedule as a function of the vaccination fraction.}
  \label{fig:per_vacc}
\end{figure}

In order to describe the quantitative effect of the different
vaccination schedules, the top panel of Fig. \ref{fig:per_vacc}
depicts the expected value of the long-term number of affected agents
as a function of the mobility parameter. The no-vaccination case,
marked with the black curve, reaches the total number of agents,
$N=1000$, with the transition at around $\ve_c\approx 0.6$. The effect
of the random vaccination scheme is analogous to that obtained by
performing a reduction of the population density with a factor
$1-f_v$, and thus the mobility parameter at the transition point will
follow the relation $\ve_c=\ve_c^{(0)}(1-f_v)^{-1/2}$, where
$\ve_c^{(0)}$ stands for the critical value of the mobility parameter
when no agents are vaccinated. Interestingly, random vaccination and
highest degree vaccination provide similar outcomes, with a slightly
higher critical value of the mobility parameter for the random
case. Indeed, this shows that highest degree vaccination is not
effective at all. Highest LBC vaccination, on the other hand, results
in a substantial reduction in the number of affected agents. Yet, the
most effective vaccination schedule is, with a large difference, the
one based on the highest BC (yellow line). This result is in agreement
with previous studies \cite{Schneider_2011,Matamalas_2018,Yang_19}.  Moreover, all schemes
reach a similar value of the total number of affected agents for large
mobility. Thus we conclude that an effective vaccination scheme will substantially increase the mobility threshold under which an epidemic
outburst will die off. Yet, the effects of vaccination can be
substantially reduced above that mobility threshold.

How effective are the vaccination schedules shifting the percolation
transition point? In order to answer that question we have traced the
percolation threshold $\ve_c$ as a function of the vaccination
fraction $f_v$ in the bottom panel of Fig. \ref{fig:per_vacc}. The
base value, for no vaccination, is $\ve_c\approx 0.6$ (as shown in
Fig. \ref{fig:clustersize}). Highest BC vaccination results in a sharp
increase of the percolation transition, reaching $\ve_c\approx 1.1$
for $\approx 10\%$ vaccination fraction. Highest LBC vaccination
performs substantially worse, but still provides a significant
improvement of the epidemiological situation for scarce
vaccines. Random and high degree vaccinations perform similarly, and
none of them are quite effective.


\section{Conclusions and further work}
\label{sec:conclusions}

In this work we have presented a very simple agent model for epidemic
expansion, the confined-SIR model, in which the effects of partial
confinements and vaccinations can be easily tested. We have developed
a theoretical framework to characterize the confined-SIR model, which
combines results from percolation theory in random geometric graphs
(RGG) with those of mean-field theory for epidemic expansion. It is
relevant to notice that our model is far too simple to lead to policy
recommendations without further insight from experts from different
fields, ranging from virology to sociology. The main shortcoming of
our model when applied to human epidemics is that human mobility is
not geographically restricted in the same way as in our model. Indeed,
even under lock-down essential workers must attend their workplaces,
which might be far away from their homes. Yet, we expect that some of
our conclusions can be interesting for researchers in epidemic
expansion and lead, in combination with other insights, to sensible
policy recommendations which will help alleviate the effects of
present and future epidemic outbursts.

Our first conclusion is that the effects of the confinement measures
vary strongly when they cross the percolation threshold. Thus, it is
	of paramount importance to design lock-down measures so that mobility
	is restricted sufficiently below the percolation threshold in order to limit the expansion of the epidemic. The
determination of the percolation threshold is not easy in practice,
but a good hint is provided by the fluctuations in the outburst
sizes. Above the transition point, most outbursts reach a huge
fraction of the population, and below it, they will only affect a few
individuals. Yet, near the transition, the number of affected agents
may vary enormously, depending on the location of the initial infected
agents. In addition, the effect of increasing the recovery probability
in our model causes a decrease of the number of final infected agents,
since it provides an efficient cutoff for the infection probability
between pairs of agents which seldom meet.

Our second conclusion is that in order to determine an efficient
vaccination schedule, disregarding other healthcare considerations,
{\em bridge} individuals should be especially targeted for vaccination,
i.e. individuals which move between different clusters. Their
immunization will lead to an effective confinement of an epidemic
outburst to its initial cluster, thus creating effective firewalls
between them. Bridge individuals can be detected via {\em
  betweenness-centrality} (BC), which is a global and computationally
expensive measure, or through easier proxies, such as the individual
{\em local betweenness-centrality} (LBC), which addresses the
question: {\em Are your friends friends among them?} Individuals whose
friends form a clique are not good candidates for vaccination, but
individuals whose friends do not know each other are. Our results in
this respect are mixed: vaccinating the highest BC agents is extremely
efficient, but finding those agents is very hard both in the
simulations and in real life. Vaccinating the highest LBC agents makes
more sense: choose the individuals whose contacts are not in contact
among themselves, such as retail salespeople. Yet, more research is
required in order to find optimal vaccination schedules.

Confined SIR models seem to be an appropriate tool in order to improve our
intuition regarding the effectiveness of different strategies to
stifle an epidemic outburst and to find relevant observables in order
to characterize the current situation, allowing us to make meaningful
predictions. For example, in this work we have emphasized the very
interesting role provided by the fluctuations in the maximal number of
infected agents. Indeed, a very promising line of research is the
statistical analysis of these fluctuations during real epidemic
outbursts, such as COVID-19. These analysis present a very interesting
challenge: fluctuations should be compared {\em ceteris paribus},
i.e. removing major differences between the different geographical
areas and times.


\begin{acknowledgments}

We would like to acknowledge J.E. Alvarellos, P. Córdoba-Torres,
R. Cuerno, E. Korutcheva and S. Ferreira for very useful
discussions. This work was funded by Instituto de Salud Carlos III
(Spain) through Grant No. COV20/01081, and the Spanish government through
Grants No. PGC2018-094763-B-I00, No. PID2019-105182GB-I00, and
No. PID2019-107514GB-100.

\end{acknowledgments}



\end{document}